\documentstyle[12pt]{article}
\begin{document}
\begin{titlepage}
\renewcommand{\thefootnote}{\fnsymbol{footnote}}

\begin{center} \LARGE
{\bf Theoretical Physics Institute}\\
{\bf University of Minnesota}
\end{center}
\begin{flushright}
TPI-MINN-94/32-T\\
UMN-TH-1313-94\\
\end{flushright}
\vspace{.3cm}
\begin{center} \LARGE
{\bf Theory of Weak Inclusive Decays and Lifetimes of Heavy
Hadrons}
\end{center}
\begin{center}
Talk at the 27th International Conference on High Energy Physics,
Glasgow, 20-27 July 1994
\end{center}
\vspace*{.3cm}
\begin{center} {\Large
Mikhail Shifman
}
\end{center}
\vspace{0.4cm}
\begin{center}
{\it  Theoretical Physics Institute, University of Minnesota,
Minneapolis, MN 55455}\\
\vspace{.3cm}
e-mail address:
 shifman@vx.cis.umn.edu

\vspace*{.4cm}

{\Large{\bf Abstract}}
\end{center}

\vspace*{.2cm}

The theory of preasymptotic effects in inclusive decays of
heavy flavors is briefly reviewed.

\end{titlepage}

\section{Introduction}

Heavy flavor hadrons $H_Q$ contain a heavy quark $Q$ plus a cloud
built from
light quarks (antiquarks) and gluons. The heavy quark
$Q$ experiences a weak transition. The nature
of this transition is of no concern to me here.
It can be a radiative transition, like $b\rightarrow
s\gamma$, semileptonic decay like $b\rightarrow
cl\nu_l$ or a non-leptonic decay to lighter quarks, e.g.
$b\rightarrow c\bar u d$.
It is assumed that at short distances the amplitude is
known from the electroweak theory. The task of the QCD-based
theory is to calculate preasymptotic effects in
the decay rate of the hadron $H_Q$ and other decay characteristics:
the energy spectra, the average invariant mass of the hadronic state
produced, etc. These effects are due to interactions with the soft
degrees of freedom in the light cloud.

The foundation of the theory was laid in the eighties \cite{1}
when it was realized that the operator product expansion \cite{2}
could be used in application to the so called transition operator
of the type
\begin{equation}
\hat T (Q\rightarrow f\rightarrow Q) = i\int d^4x
\{ {\cal L}_W(x), {\cal L}^\dagger_ W(0)\}_T ,
\label{trans}
\end{equation}
where ${\cal L}_W$ is the short-distance weak Lagrangian
governing the transition $Q\rightarrow f$ under consideration.
The momentum operator ${\cal P}_\mu$ of the heavy quark $Q$ is
written as a sum of two terms, ${\cal P}_\mu = m_Qv_\mu +
\pi_\mu$
where $v_\mu$ is the
four-velocity of the heavy {\em hadron} $H_Q$, $\pi_\mu $
is the residual momentum operator,
responsible for the interaction with the ``background" gluon field in
the light cloud. The large mechanical part $m_Qv_\mu$ in the
heavy quark momentum guarantees that the transition operator
(\ref{trans}) can be found as a sum of {\em local} operators
(with some reservations to be discussed below). These operators are
ordered according to their dimensions. At the level of
$1/m_Q^2$ we have only two operators; extra four-fermion operators
are added at the level of $1/m_Q^3$. A few of these were calculated
8 years ago \cite{1}.

At the next stage each term in the expansion must be averaged over
the hadronic state $H_Q$. At this stage the bound state dynamics
is accounted for.

After the initial excitement the OPE-based theory of the inclusive
heavy flavor decays was in a rather dormant state until recently.
The revival it experiences now is due to a combination of several
factors. First, a very concise and convenient language was created,
the heavy quark effective theory (HQET) \cite{HQET}. Calculations of
the non-perturbative effects were translated in this language and
developed in Refs. \cite{Chay,Bigi,Blok}. Second, all relevant
operators appearing at the level up to $1/m_Q^3$ were catalogued
and our understanding of their matrix elements (I mean the
numerical values) was significantly advanced. Finally, the issue of
convergence of the non-perturbative series was clarified.
In the
beauty family one expects that the first two or three terms in the
expansion ensure reasonable accuracy of the predicted lifetimes.
As for  the charmed quark we will see that  it is, perhaps,
too
light for duality to set in. Since OPE is
used in the Minkowski domain the validity of duality is crucial
for the whole approach. Theoretically the onset of duality is
correlated
with the behavior of high-order terms in the non-perturbative
series.

\section{Master equation}

The OPE-based approach is applicable in a very wide range of
problems. Here we will concentrate on the total inclusive widths.
Generically the $m_Q^{-1}$ expansion for the width has the form
(for definiteness I will speak about the beauty family)
$$
\Gamma (H_b \rightarrow f)=\frac{G_F^2 m_b^5}{192\pi^3}
|{\rm CKM}|^2\times
$$
$$
\left\{
c_3 (f) \frac{\langle H_b|\bar bb|H_b\rangle}{2M_{H_B}} +
\frac{c_5 (f)}{m_b^2}
\frac{\langle H_b|\bar b(i/2)\sigma Gb|H_b\rangle}{2M_{H_B}}+
\right.
$$
\begin{equation}
\left.
 \sum_i
\frac{c_6^{(i)} (f)}{m_b^3}
\frac{\langle H_b|(\bar b\Gamma_i q)(\bar q\Gamma_i
b)|H_b\rangle}{2M_{H_B}} +{\cal O}(m_b^{-4})\right\} \, .
\label{master}
\end{equation}
The coefficient functions $c_i (f)$ depend on the particular inclusive
transition considered and are calculable. They are determined by
short-distance QCD provided that the energy release is large enough.
On the other hand, the matrix elements on the {\em rhs}
describe the response of the soft degrees of freedom on the
instantaneous perturbation, the $b$-quark decay. These quantities
are
essentially non-perturbative. But they are universal, and, as seen
from Eq.
(\ref{master}), there are only a few of them.

The matrix element of the chromomagnetic operator $\sigma G$ is
expressible
through spin splittings, say $M_{B^*} - M_B$. The four-fermion
operators of dimension 6 can be evaluated, in the case of mesons,
within factorization. For baryons a reliable  calculation of the
corresponding matrix elements is a  problem
essentially unsolved so far. As for the scalar density, $\bar b b$, it is
this term that exactly reproduces the parton model (asymptotic)
result, plus preasymptotic corrections.
This operator also can be written as an expansion,
$$
\bar b b = \bar b\gamma_0 b - \frac{1}{2m_Q^2}
\bar b ({\vec\pi}^2- (i/2)\sigma G)b +
$$
\begin{equation}
\frac{1}{4m_Q^3}g^2 \bar b\gamma_0 T^a b
\sum_q \bar q\gamma_0T^a q +
{\cal O}(1/m_b^4)
\end{equation}
where the sum runs over the light quarks. A new operator appearing
here is $\bar b{\vec\pi}^2 b$, the square of the spatial momentum of
the b-quark. The matrix element of this operator in the $B$
meson can be limited from below, for a detailed discussion
see Ref. \cite{BSUV}. The average spatial momentum turns out to be
surprisingly large, larger than 0.6 GeV! The QCD sum rule calculations
\cite{Ball} yield even a larger value, $\sim$ 0.7 GeV. The expectation
value of
${\vec\pi}^2$ in baryons is expected to be close to that in mesons.

Time/space limitations do not allow me to go into further details.
Let me point out only the most remarkable features of the overall
picture.

(i)  The total rates do not contain  non-perturbative corrections
of order $1/m_b$, the so called CGG/BUV theorem \cite{Chay,Bigi}.
The corrections start at the level $1/m_b^2$. This sets the scale of
preasymptotic effects in the beauty family at the level of several per
cent since $\langle B|\bar b i\sigma G b|B\rangle/2m_b^3\sim 0.03$.
In particular, deviations of the lifetime ratios from unity are
expected to be
of this order of magnitude.  At the level $1/m_b^2$ all $B$ mesons
have the
same lifetimes (disregarding some small $SU(3)_{\rm fl}$ breaking
effects).

(ii) The difference in the lifetimes of baryons and mesons
is due to the fact that the expectation values of the operators in Eq.
(\ref{master}) are different for mesons and baryons. This difference
arises at the level $1/m_b^2$.

(iii)
Four-fermion operators of dimension 6 produce effects formally
scaling
like $1/m_b^3$, although numerically they seem to be enhanced
since the corresponding coefficients have one loop less and,
additionally,  a key constant
 $f_B$ turns out to be rather large. This enhancement may lead
to the fact that dimension 5 and 6 operators are competitive in the
beauty family. The four-fermion operators shift the lifetimes of
mesons versus baryons and split the meson lifetimes from each
other.

(iv)   Situation with  $B_s$ is exceptional. The lifetime difference
between $B_{s,{\rm short}}$ and $B_{s,{\rm long}}$ is due to a
mechanism not
exhibited in Eq. (\ref{master}), namely $B-\bar B$ oscillations. The
corresponding estimates were done in Ref. \cite{Khoze}.

\section{Phenomenological implications}

Assembling all theoretical elements discussed above (and those
which are discussed in the original literature) we arrive at the
following pattern. The lifetime of a charged $B$ meson is predicted to
exceed that of a neutral $B$ meson,
\begin{equation}
\frac{\tau (B^-)}{\tau (B_d)}-1 \approx 0.05(f_B/200\mbox{MeV})^2
\sim 0.05 .
\end{equation}
At this level it is expected that $\bar\tau (B_d)\approx \bar\tau
(B_s)$
where $\bar\tau$  denotes the average lifetime of the two mass
eigenstates in the $B^0$--$\bar B^0$ system. It is curious that
$B_s$ oscillations will seemingly produce the largest
lifetime difference,
\begin{equation}
\frac{\Delta\Gamma (B_s)}{\bar\Gamma (B_s)}
\approx 0.18(f_B/200\mbox{MeV})^2\sim 0.18\, .
\end{equation}
The baryon matrix elements are always most difficult for
consistent analysis; therefore, the baryon-to-meson lifetime ratios
should be taken with caution. Still, plausible estimates indicate that
one can expect $\tau (\Lambda_b)/\tau (B_d)\sim 0.9$.

\section{A grain of salt: ${\rm Br}_{\rm sl} (B)$}

The theory of preasymptotic effects which I have just sketched,
being
applied to the problem of the semileptonic branching ratio in the $B$
mesons, leads to a paradox. In this case the heavy quark expansion
can be readily carried out up to terms of order $1/m_b^3$.
One obtains a formula very similar to Eq. (\ref{master}), with the
same structure and the same operators \cite{BBSV}. The leading
non-perturbative correction ${\cal O}(m_b^{-2})$ tends to diminish
the branching ratio while the term ${\cal O}(m_b^{-3})$ tends
to increase it. Both effects, however, are far too small to
produce a noticeable impact on the branching ratio. At best they shift
the prediction for the branching ratio by $0.5\%$ or less.
Thus we are forced to conclude that the prediction for
${\rm Br}_{\rm sl} (B)$ is controlled by perturbative QCD. People
believe that perturbative QCD  typically yields ${\rm Br}_{\rm sl}
(B)\approx 13\%$; twisting arms allows one to go down to
$12.5\%$ \cite{Altarelli}. At the same time experimentalists, both
CLEO and ARGUS, seem to be firm in their conclusion that
${\rm Br}_{\rm sl} (B)< 11\%$. A natural question is what went
wrong?

I leave aside the possibility that the experimental numbers are
wrong. There is no visible loophole in the OPE-based theory of
preasymptotic effects either. Then the remaining logical options are
as follows: (i) something is missing in the perturbative analysis;
(ii) new physics shows up in the $B$ meson decays. Both options
must be investigated. In a recent paper \cite{Ball2} a new
contribution in the perturbative calculation is identified, not included
in the analysis of Ref. \cite{Altarelli}, which seemingly works in the
right direction -- diminishes ${\rm Br}_{\rm sl} (B)_{\rm pert}$ by
$\sim 0.5\%$. It remains to be seen whether the perturbative
number can reach the $11\%$ mark under  realistic choice
of relevant theoretical  parameters (the quark masses, $\alpha_s$,
etc.). (Let me make a side remark: I do not believe
that $\alpha_s (M_Z)$ can be as large as 0.126, as is allegedly
implied by the so called
global fits at the $Z$ peak at present. A wealth of low-energy data
point to a significantly lower value of $\alpha_s$, something like
0.114 or even lower.  In terms of $\Lambda_{\rm QCD}$ the
difference is drastic. I urge to take this discrepancy very seriously.)

\section{The family of charm}

It might seem to be a trivial exercise to substitute $m_b$ by $m_c$
in the master equation. Yes, technically this is easy, and
formally all $1/m_c^2$ and $1/m_c^3$ corrections have been
written down and classified. They are much larger, of course,
than in the beauty family; typically of order of 0.5. I refer those
interested to a very detailed recent update \cite{BS}. Qualitatively
the pattern of the lifetimes in the charm family emerging in the
heavy quark expansion agrees with experiment. Namely, those
particles that live longer are predicted to live longer, etc. However,
quantitatively
the ${\cal O}(m_c^{-2})$ and ${\cal O}(m_c^{-3})$ preasymptotic
terms are smaller than what one needs in order to reproduce, say,
$\tau (D^+)/\tau (\Xi_c^0)_{\rm exp}\sim 12$. Since arithmetically
the
calculation is certainly correct one may start suspecting that
something went wrong in the basics.

The operator product expansion, the foundation of the whole
approach, is a well defined procedure in the Euclidean domain.
A specific feature of the transition operator (\ref{trans})
is its essentially Minkowski character. Therefore, in justifying the
short-distance calculation of the coefficient functions one must
always
keep in mind a kind of analytic continuation, through a dispersion
relation. Thus, strictly speaking, theoretical predictions for
$c_i (f)$  in Eq. (\ref{master}) refer to quantities
integrated over energy in some energy range.

If the
energy release is large
enough so that duality is valid and the integrand is smooth,
this smearing is unimportant; one can predict the coefficient
functions for the given energy release, locally. It is always tacitly
assumed
that this is the case. The onset of duality is governed by  exponential
terms, not visible to any finite order in $1/m_Q$ expansion.
Due to this reason the onset must be abrupt.

We are inclined to think that the $c$ quark is not heavy enough
to warrant duality. The strongest argument comes from
consideration of $\Gamma_{\rm sl}(D)$.
Indeed, with the reasonable value of $m_c$ ($m_c(m_c)\sim 1.3$
GeV) the parton-model
prediction is close to the experimental number. The first
perturbative correction is negative \cite{Altarelli} and the second
seems to be negative as well \cite{Luke}. The non-perturbative
corrections follow the same pattern. The leading $1/m_c^2$ term is
known from Ref. \cite{Bigi} while the $1/m_c^3$ correction
has been estimated recently \cite{BDS}; both are negative. The
combined effect of the leading corrections amounts to reducing
$\Gamma_{\rm sl}(D)$
by  $50\%$, and the next-to-leading terms only worsen the situation!

This work was supported in part by DOE under the grant number
DE-FG02-94ER40823.

\end{document}